\documentclass[fleqn,10pt]{wlscirep}
\title{Hope4Genes: a Hopfield-like class prediction algorithm for transcriptomic data}

\author[1,2,*]{Laura Cantini}
\author[3]{Michele Caselle}
\affil[1]{PhD in Complex Systems for Life Sciences,University of Torino, Torino, Italy}
\affil[2]{Institut Curie, PSL Research University, INSERM U900, Mines ParisTech, Paris, France}
\affil[3]{Università degli Studi di Torino, Department of Physics and INFN, via P.Giuria 1, I-10125 Turin,Italy}

\affil[*]{laura.cantini@curie.fr}


\begin{abstract}
After its  introduction in 1982, the Hopfield model has been extensively applied for classification and pattern recognition. Recently, its great potential in gene expression patterns retrieval has also been shown. Following this line, we develop Hope4Genes a single-sample class prediction algorithm based on a Hopfield-like model. Differently from previous works, we here tested the performances of the algorithm for class prediction, a task of fundamental importance for precision medicine and therapeutic decision-making. Hope4Genes proved better performances than the state-of-art methodologies in the field independently of the size of the input dataset, its profiling platform, the number of classes and the typical class-imbalance present in biological data. Our results provide encoraging evidence that the Hopfield model, together with the use of its energy for the estimation of the false discoveries, is a particularly promising tool for precision medicine.\\
\\ 
\end{abstract}
\begin{document}

\flushbottom
\maketitle
%
%
\thispagestyle{empty}

\section*{Introduction}

The Hopfield model is the most popular and paradigmatic example of attractor neural network, i.e. a network whose spins evolve towards stored attractor patterns \cite{Hopfield82}. From its introduction in 1982, the Hopfield model has been widely applied in computer science as associative memory to store a number of patterns and retrieve them from noisy/corrupted inputs. Its application for classification and pattern recognition spans numerous fields of research, from economy to medicine and social sciences \cite{Yalcinoz97,Lin96}. In the past few years, thanks to the increasing availability of high-throughput data, the Hopfield network has been also applied to describe the emergence of gene expression patterns \cite{Maetschke14,taherian2017} or, even more ambitiously, to  model tissue differentiation \cite{Szedlak14,Anafi10,Lang14} and to estimate the pseudotime from single-cell RNA-seq data \cite{guo2017,jang2017,fard2016,szedlak2017}. These last studies proved that the memory retrieval ability, characteristic of the Hopfield model, has a great potential for the analysis of transcriptomic data.\\ 
\\
One of the major issues of personalized medicine is the classification of patients affected by a given disease into groups homogeneous with respect to molecular, prognostic or therapeutic features \cite{van2008enabling}. The five molecular subtypes of breast cancer (Luminal A, Luminal B, HER2-enriched, Basal-like and Normal-like), identified and intensively studied in the last 10 years, represent a typical example \cite{perou2000,sorlie2001gene}. Multiple approaches have been proposed for cancer subtyping, most of them based on: matrix factorization \cite{sadanandam2013}, clustering \cite{marisa2013}, networks \cite{wang2014} and deep learning \cite{guo2018b,guinney2015}. All these algorithms are aimed at performing an unsupervised classification of a set of patients and at identifying a gene signature correlated with the identified classes. Once the subtypes have been identified, their usage for therapeutic decision-making on a new patient, requires to solve a class prediction problem. Class prediction requires in input a signature, i.e. a set of genes coherently over- or under-expressed in patients with a certain phenotype of interest, and it is aimed at associating each sample of the input dataset to a class (e.g. poor vs good prognosis) based on the coordinate behavior of the genes belonging to the signature. \\
\\
In the last years, transcriptomic signatures able to differentiate distinct subtypes of tumors, to identify important cellular responses, to predict clinical outcomes and to model the activation of signaling pathways have been extensively developed \cite{Liu07,Ferrando02,Boersma08}. Consequently to the spread in gene signatures production, many class prediction algorithms have been introduced. Among them Nearest Template Prediction (NTP), based on associating a sample to the class centroid at lower distance, was found to outperform previous methods and it represents the state-of-art in the field \cite{Hoshida10}. The main advantages of NTP with respect to previously designed algorithms, such as those based on supervised learning approaches \cite{guo2018bcdforest}, are represented by: (i) the absence of a  training phase and the consequent risk of overfitting; (ii) the possibility to classify one single sample per time, fundamental for an effective implementation of the tool in the clinics and (iii) the implementation of a False Discovery Rate (FDR) to asses the prediction confidence.\\
\\
Due to the complexity of cancer, the high-dimensionality of trascriptomic data and the class-imbalance typical of these classification problems, the overall performances of most current methods still need to be further improved and more accurate and robust methods need to be developed.\\
\\
Inspired by the promising results obtained by the Hopfiled model in the analysis transcriptomic data, we here developed Hope4Genes a new algorithm for class prediction based on a Hopfield-like model. Hope4Genes, available at \url{https://github.com/lcan88/Hope4Genes}, preserves the three advantages implemented in NTP (absence of training phase, single-sample classification and FDR estimation). Moreover, it associates to each sample a secondary class, useful to better characterize those $5-10\%$ of samples that generally result to be unclassified by existing algorithms. Once tested on the same data to which NTP was originally applied, Hope4Genes obtained competing results, proving that the application of the Hopfield model for class prediction of transcriptomic data, interesting in itself, has also a great potential in precision medicine.

\section*{Results}

\subsection*{Hope4Gene a class prediction algorithm based on the Hopfield model}

\begin{figure}[ht]
\centering
\includegraphics[width=\linewidth]{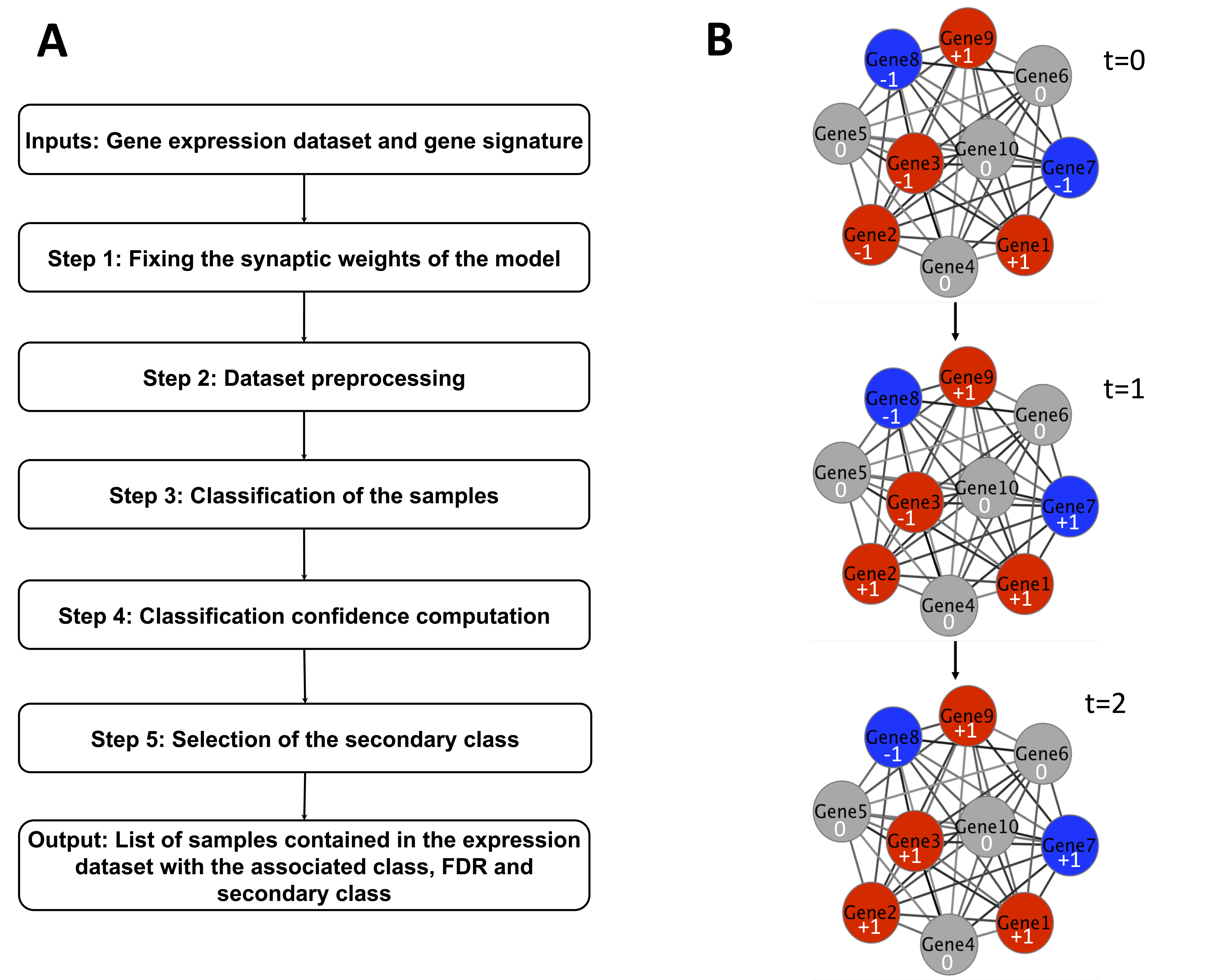}
\caption{A. Schematic representation of Hope4Genes algorithmic steps; B. Example of application of Hope4Genes. Considering a two-classes (1 and 2) classification problem, the complete network having as nodes the genes belonging to both the signatures of class 1 and 2 is reconstructed. The weights of the network's links are set using the Hebb's rule to have the templates of classes 1 and 2 stored into the model. Genes belonging to the signature of class 1 are denoted in red, while the genes of class 2 are denoted in blue. Considering that, for example, class 1 has more than $\Gamma$ genes, the grey nodes denote those genes of the signature of class 1 that were not selected due to the $\Gamma$ thresholding. Once the network is reconstructed, to classify a sample, we first discretize the expression values of the signature genes in the sample into $\{+1,-1\}$ and we assign each value to the corresponding node (t=0 in the figure). We then let the model evolve (from t=0 until t=2) when only one of the two classes (class 1) will have all values $+1$. We finally assign the sample to the class of convergence (class 1).}
\label{fig:pipeline}
\end{figure}

We here develop Hope4Genes, available at \url{https://github.com/lcan88/Hope4Genes}, a single-sample class prediction algorithm for transcriptomic data (see Figure \ref{fig:pipeline}). Given a gene expression matrix, a set of classes in which the user wants to classify the samples together with a signature, i.e. a set of genes whose expression is correlated to a class, the general idea behind Hope4Genes is to inizialize the nodes of the Hopfield network according to the binarized expression values of each sample and let the model evolve until it reaches one of the class templates stored into the model (see Methods for a detailed explanation of the Hope4Genes algorithm). Indeed the Hopfield network consists of a set of neurons connected through symmetric bidirectional links and having two possible states: firing ($+1$), or silent ($-1$). A set of patterns can be stored in a Hopfield model by setting the weights of its links based on the Hebb rule, that assures to have the stored patterns as fixed points of the dynamics, or equivalently as minima of the energy associated to the model (see Methods). An example of application of Hope4Genes is reported in Figure \ref{fig:pipeline}B.\\  
\\ 
A straightforward application of the Hopfield model to class prediction has two main limitations: (i) gene signatures are typically mutually non-overlapping partitions of the gene set, thus class templates are strongly correlated and (ii) gene signatures are usually strongly unbalanced, in fact their sizes usually show a wide variability, with classes represented by hundreds of genes and classes composed by few dozens of genes. The consequence of (i) is that the Hopfield model can converge to spurious memories. To overcome this problem, we modified the classical model proposing an asymmetric, diluted version of the Hopfield model (see Methods). To face problem (ii), we performed a signature reduction combined with a resampling procedure and we analytically derived the optimal threshold ($\Gamma$) for the size of the signature reduction (see Methods). For these reasons we refer to the algorithm here proposed as a Hopfield-like model. An example of the final structure of the model is reported in Figure \ref{fig:pipeline} A. A False Discovery Rate (FDR), estimated based on the energy landscape of the Hopfield model, is also associated to each sample (see Methods). Finally, when more than two classes are considered, Hope4Genes allows to obtain the second best prediction and thus characterize those samples that are a mixture of multiple classes.

\subsection*{Hope4Genes classification performances}
We tested Hope4Genes on five examples, chosen to explore the performances of the algorithm when the sample size, the number of classes and the profiling platform of the input data change. We thus considered multi-class and two-class classifications, data profiled with three different microarray platforms (Affimetrix, Illumina and Agilent) and RNAseq, data composed of different sample sizes (from 12 to 397). The majority of the considered tests have been obtained from the original publication of Nearest Template Prediction (NTP) \cite{Hoshida10}, representing the current state-of-art in the field of class prediction for transcriptomic data \cite{sadanandam2013,isella2015,cantini2015}. We considered in particular:  
\begin{itemize}
\item \textbf{Example 1:} Prediction of Acute Lymophoblastic Leukemia (ALL) and Acute Myeloid Leukemia (AML). 35 leukemia samples to be classified into AML and ALL based on the gene signature defined in \cite{Golub99}.
\item \textbf{Example 2:} Cross-platform prediction of Estrogen Receptor (ER) positivity in breast cancer. 49 breast cancer samples from \cite{West01} to be classified according to the gene signature obtained in \cite{Hoshida10}.
\item \textbf{Example 3:} Cross-species prediction of liver cirrhosis between human and rat. 12 liver cirrhosis rat samples are classified through a signature generated on human samples affected by the same pathology in \cite{Wurmbach07}. 
\item \textbf{Example 4:} Prediction of multiple tissue types. A dataset containing 52 samples collected from four different tissues: breast, prostate, lung and colon is classified based on the signature defined in \cite{Su02}. 
\item \textbf{Example 5:} Prediction of breast cancer molecular subtypes: Basal-like, HER2, Luminal A, Luminal B, and Normal. This last example is the one representing a real-world clinical application. To study how the preformances of the algorithm are influenced by the size of the input dataset, its profiling platform and the typical class-imbalance present in biological data we divided the example in four subcases obtained by varying the transcriptomic dataset and the considered subtypes (from 3 to 5):
\begin{itemize}
\item \textbf{Example 5.1}: a dataset of 198 samples \cite{Weigelt10} is classified into 5 subtypes: Luminal A, Luminal B, HER2, Basal and Normal.
\item \textbf{Example 5.2}: we classified a dataset of 286 samples \cite{Desmedt10} in 4 subtypes:  Luminal A, Luminal B, HER2 and Basal.
\item \textbf{Example 5.3}: 53 samples of \cite{Wang05} are classified into 3 subtypes: Basal, Luminal A and Luminal B. 
\item \textbf{Example 5.4}: BRCA Level 3 RNA-seq expression profiles downloaded from TCGA in January 2016 and consisting of 397 samples are classified into 4 subtypes:  Luminal A, Luminal B, HER2 and Basal. This last test was not present in \cite{Hoshida10} and it has been added in order to compare the classification performances also on RNAseq data.
\end{itemize}
The used gene signature has been constructed in \cite{Hoshida10}. 
\\
\end{itemize}

NTP was also applied to the same examples and the performances of the two algorithms are compared in the following sections. The datasets, class labels and signatures used for the comparison were downloaded from \url{http://www.broadinstitute.org/cgi-bin/cancer/datasets.cgi}. The size of the signatures ranges from 19 to 956. A detailed summary is reported in Supplementary Table 1.\\
NTP was applied with standard parameters (cosine distance and 1000 permutations) through the GenePattern analysis toolkit (\url{www.broadintitute.org/genepattern}) \cite{GP}. \\
In line with \cite{Hoshida10}, additional prediction analyses were performed with standard machine learning methods: Classification and Regression Trees (CART), Weighted Voting (WV), Support Vector Machine (SVM) and k-Nearest Neighbor (k-NN). All the algorithms have been applied with standard parameters through the GenePattern analysis toolkit (\url{www.broadintitute.org/genepattern}) \cite{GP}. Differently from NTP and Hope4Genes these alternative approaches involve a training and test phases. Training datasets were downloaded from \url{http://www.broadinstitute.org/cgi-bin/cancer/datasets.cgi}. For Example 5.4 the Metabric dataset, available at \url{https://www.synapse.org/#!Synapse:syn1688369} was employed as training set. For multi-class classifications (Examples 2-3 and 5) the expression levels were z-score transformed at gene-level to adjust the range of expression of each gene between training and test datasets. 

\subsubsection*{Hope4Genes vs. NTP: comparison without taking into account the FDR}
\begin{figure}[ht]
\centering
\includegraphics[width=\linewidth]{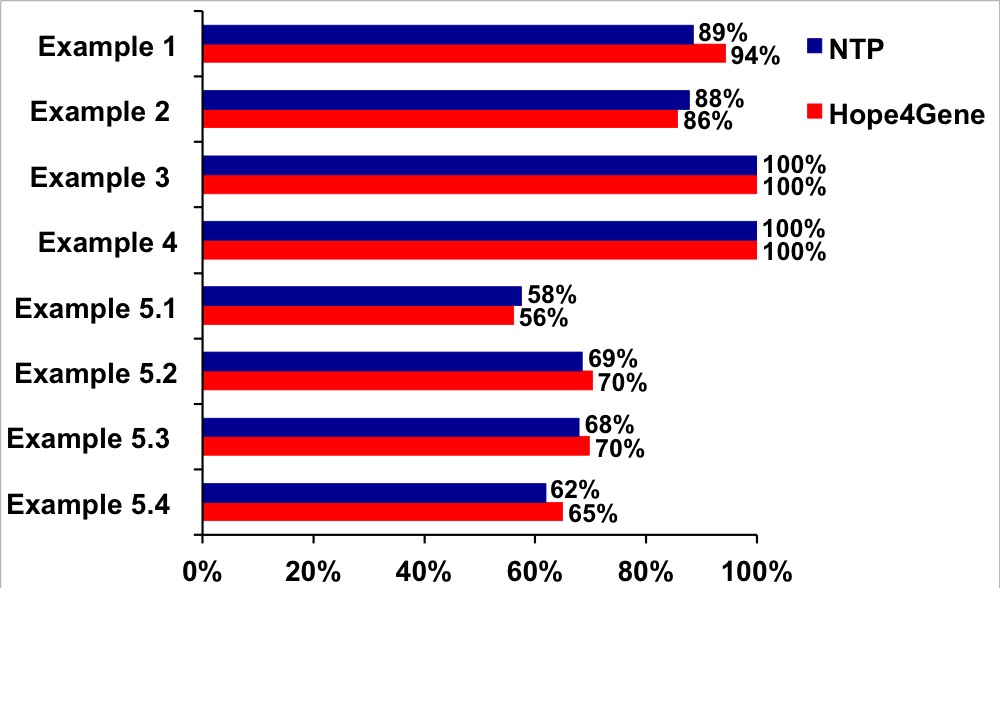}
\caption{Hope4Genes vs. NTP classification performances without FDR. Histograms reporting the percentage of correctly classified samples in the seven Examples according to Hope4Genes (red) and NTP (blue).}
\label{fig:comp_noFDR}
\end{figure}

First, we compared the classifications provided in the five tests by Hope4Genes against those of NTP without taking into account the FDR. As shown in Figure \ref{fig:comp_noFDR}, two tests (examples 3 and 4) were perfectly classified by both the algorithms. In four of the remaining tests (examples 1, 5.2, 5.3 and 5.4), Hope4Genes performed better than NTP, while in the remaining two tests (examples 2 and 5.1) NTP obtained better results. Overall,  when the FDR is not taken into account, Hope4Genes tends to perform only slightly better than NTP with an overlap in the predictions between the two of $87\%$ on average.\\

\subsubsection*{Hope4Genes vs. NTP: comparison taking into account the FDR}
\begin{figure}[ht]
\centering
\includegraphics[width=\linewidth]{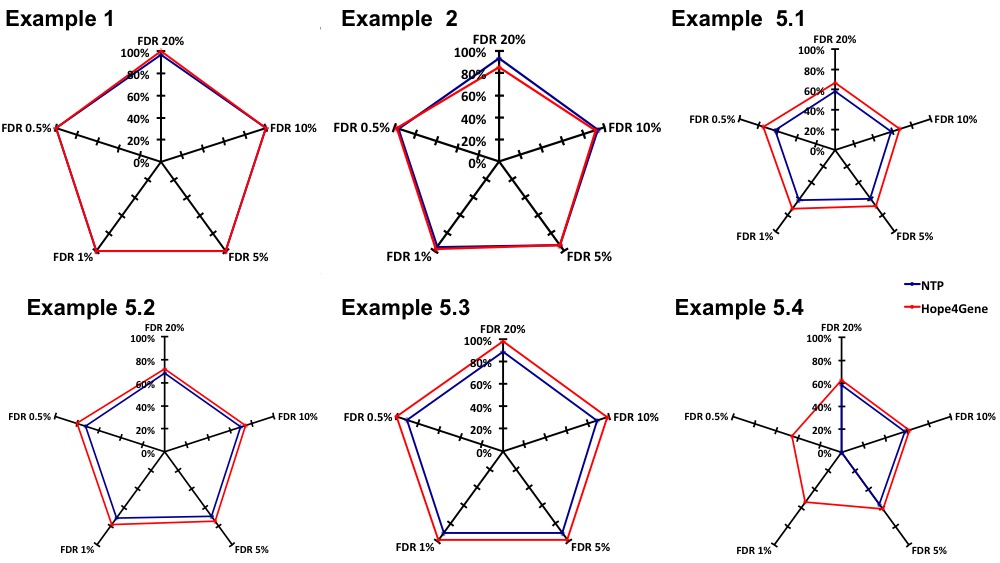}
\caption{Hope4Genes vs. NTP classification performances with FDR. Radar plots reporting the percentage of correctly classified samples in the five remaining Examples according to Hope4Genes (red) and NTP (blue).}
\label{fig:comp_FDR}
\end{figure}

We then compared the performances of the two algorithms considering only significant (sample,class) associations. The FDR thresholds that we used are standard ones ($20\%$, $10\%$, $5\%$ and $1\%$) plus the more stringent $0.5\%$. From this comparison we excluded examples 3 and 4 for which a perfect classification was already obtained without taking into account the FDR. The results obtained in the remaining 6 cases are summarized in the radar plots of Figure \ref{fig:comp_FDR}, where each radius is associated to a different FDR threshold. As shown in the figure, in the four tests in which Hope4Genes had been already found to outperform NTP (examples 1, 5.2, 5.3 and 5.4), the results are here further confirmed.\\ 
\\
More interesting are the results obtained in those two tests (examples 2 and 5.1) in which NTP had been found above to perform slightly better than Hope4Genes. Indeed, in example 2, while for high FDR thresholds ($20\%$ and $10\%$) NTP performs better than Hope4Genes, for an FDR of $5\%$ Hope4Genes reaches the same classification performances of NTP and it finally outperforms NTP, for FDR values of $1\%$ and $0.5\%$. In example 5.1, the use of the FDR completely overcomes the results obtained above. Indeed Hope4Genes results to perform better than NTP for all the five FDR thresholds, reaching a percentage of correctly classified samples greater than that obtained by NTP of $8-12\%$. These results confirm that the use of an Hopfield-like model and of its energy for the estimation of the FDR leads to an increase in classification specificity in respect to NTP.\\
A comprehensive comparison of Hope4Genes and NTP (with and without FDR) together with more standard machine learning methods: CART, WV, SVM and k-NN is reported in Supplementary Table 2.
\\
Finally, Hope4Genes also associates to each sample a secondary class giving a more exhaustive characterization of the expression patterns characteristic of each sample. Existing classification algorithms find a $5-10\%$ of samples with no clear association to any class ( i.e. unclassified). This frequently does not result from a limitation of the algorithm, but because at the molecular level these samples are a mixture of multiple classes. This holds true in all those contexts, as cancer subtyping, in which an evident distance among the classes does not exist. The secondary class provided by Hope4Genes thus helps to better characterize the molecular features of the samples under analysis. 

\section*{Discussion}
In recent years, the classification of cancer samples into subtypes has become one of the major issues of precision medicine. This is even stronger for cancer whose response to treatment regimes and disease course and severity vary going from one patient to another, even if both affected by the same tumor type. To achieve the best therapeutic performances, cancer patients need to be classified into subpopulations characterized by a common prognosis or response to treatment (“cancer subtypes”). The association of a new patient to predefined subtypes, also known as class prediction, is thus vital to precision medicine. We here proposed Hope4Genes aimed at achieving this goal using a Hopfield-like model.\\
\\
We tested the performances of Hope4Genes in multiple test cases, exploring datasets profiled with different platforms, of different sample-sizes and multi-class and two-class classifications. In these tests Hope4Genes was found to obtain reliable results and to outperform the current state-of-art in the field (NTP), when the classification FDR is taken into account. These results prove that our implemented FDR computation based on the energy associated to the Hopfield model is able to better capture classification uncertainty than the NTP one based on signature genes resampling. Moreover, our algorithm also provides an extra information in respect to NTP, that is the secondary class, useful to characterize unclassified samples. Taken together the obtained results prove that the application of the Hopfield model for class prediction of gene expression data, interesting in itself, is also particularly well performing. The applications of Hope4Genes here proposed do not show the full potential of this model. Indeed the performances of the Hopfield model are maximized when dealing with signatures that are overlapping and redundant, as those required, for example, in the reconstruction of differentiation trajectories from single-cell RNA-seq data.\\
\\
In the last years, the number of available gene expression signatures is getting larger \cite{cantini2017}. Fast and precise class prediction algorithms, as the one here proposed, will be more and more fundamental to fully exploit the power of the available gene signatures. The use of class prediction algorithms will be required both in the context of genomic data analysis and in everyday clinical practice. Hope4Genes has thus the potential to impact not only computational biology, but also clinical decision-making, becoming a standard procedure applied by oncologists who need to choose the best performing therapy for a given patient.\\ 
\\
In this work, we only focused on single-sample class prediction for transcriptomic data, because mostly transcriptome is currently used to classify samples. However Hope4Genes can be in principle employed also to classify methylation and copy number data when signatures composed of epigenetic markers or chromosomes segments are available \cite{ernst2011}. Our approach could be thus naturally extended in the future for the integrative class prediction of multiple omics data. At the same time, given the results obtained by the Hopfield model in the context of single-cell pseudotime reconstruction, once stable signatures for the classification of single-cell RNA-seq data will be available, our algorithm could be also extended for the classification of these data. 

\section*{Methods}

\subsection*{The Hopfield model}
The Hopfield model is a fully connected neural network able to recall stored memories starting from a noisy or distorted input. The Hopfield network consists of $N$ neurons connected through symmetric bidirectional links. We will denote the activity of neuron $i$ with the binary variable $x_i \in\{+1,-1\}$, corresponding to the two possible neurons states: firing ($+1$), or silent ($-1$). As a consequence, the state of the network can be represented as a binary vector $x=(x_1 \ldots x_N)$ whose i-th component, $x_i$ represents the state of neuron $i$. The neurons interactions are encoded in the connection matrix, a $N \times N$ real symmetric matrix without self-interaction terms whose entries $W_{ij}$ define the weight of the connection between neuron $i$ and $j$. \\
Denoting $x^{t}_i$ the state of the neuron $i$ at time $t$, the time evolution of the network is determined by the following updating rule: 
\begin{equation}\label{agg}
x^{t+1}_i=sign\bigg(\sum\limits_{j=1}^N W_{ij} x^{t}_j\bigg).
\end{equation}
The dynamics described in (\ref{agg}) can be performed either synchronously or asynchronously. In the first case, at each time $t$, the state of all the neurons is updated simultaneously. Contrary, in asynchronous updating, at each time one node chosen at random changes its state depending on the state of its neighbours. We here chose to use asynchronous updating which has better convergence properties since it avoids spurious cycles.\\
The Hopfield network is generally used for pattern storage and recovery.
In order to be stored in the network a set of $p$ patterns $\xi^{\mu }=(\xi_1^{\mu } \ldots \xi_N^{\mu })$, with $\mu=1,\ldots,p$ should be stable fixed points of the dynamics (\ref{agg}).  
This can be achieved in two steps.
First, it can be shown that asynchronous updating and  a symmetric non-negative diagonal $W$ are sufficient conditions for the convergence of the recursion (\ref{agg}) to its stable 
states\cite{Hopfield82,Bruck88}. The proof of this statement is generally achieved by introducing the energy function:
\begin{equation}\label{energy}
E = -\frac{1}{2} \sum\limits_{i,j=1}^N  W_{ij} x_i x_j,
\end{equation}
and proving that this is a Lyapunov function for the system. If this is verified, the energy (\ref{energy}) does not increase at each state transition and the system evolution ends in a local minimum of the energy.\\
Second, it can be shown that with the following choice for the connection matrix:
\begin{equation}\label{hebb}
W_{ij}=\frac{1}{N}\sum\limits_{\mu=1}^p \xi^{\mu}_i \xi^{\mu }_j,
\end{equation}
known as Hebb rule, if $p$ is not too large, the patterns $\xi^{\mu }=(\xi_1^{\mu } \ldots \xi_N^{\mu })$, $\mu=1,\ldots,p$ are among the stable fixed points of the dynamics defined by eq.(\ref{agg}).\\
The number $p$ of patterns that can be stored in the network with this rule is finite and proportional to the number of neurons $N$. 
There is a critical value $p_c$, called storage capacity, such that only if the number of patterns is smaller than $p_c$ they can 
be recovered by the model. In particular it can be shown that for  random uncorrelated memories $p_c/N \sim 0.138$ \cite{Amit85}.
\\ 
Given all the characteristics described above, the Hopfield model is perfectly suited for class prediction of gene expression data.

\subsection*{Hope4Genes}
  
Samples classification is classically performed starting from an expression matrix, a predefined number of classes $p$ and a gene signature associated to each class. A signature is a set of genes whose expression was found to be representative of a given class in an independent dataset. The aim of class prediction is to associate a sample to a class using only the expression pattern of the signature genes in the analyzed sample. This problem can be formalized through an Hopfiled model considering a sample as the noisy/distorted version of one of the $p$ "class templates". The template $T_{\mu}$ for class $\mu \in \{1\ldots p\}$ is a binary vector, whose $i$-th component is $+1$ if the $i$-th gene belongs to the signature of class $\mu$ and $-1$ otherwise. Its size, in concordance with \cite{Hoshida10}, is:
\begin{equation}\label{size}
N= \sum^{p}_{\mu=1} N_\mu,
\end{equation}
where $N_{\mu}$ denotes the size of the $\mu$-th gene signature. In our application, the number of memories $p$ is equal to the number of subtypes in which the expression data need to be classified and it is thus much lower than the number of network nodes $N$, corresponding to the quantity in formula (\ref{size}). However, the hypothesis of having uncorrelated memories does not hold true in our application. In fact, gene signatures are typically mutually non-overlapping partitions of the gene set, thus class templates are strongly correlated. Our Hopfield model can thus converge to spurious solutions. At the same time, gene signatures are usually strongly unbalanced, in fact their size usually shows a wide variability with classes represented by hundreds of genes and classes composed by few dozens of genes.\\
\\
To overcome both these problems, we modified the classical model proposing an {\it asymmetric, diluted version of the Hopfield model} which is able to classify samples according to a given gene signature better than existing state of art algorithms. Our proposal has two ingredients:
\begin{enumerate} 
\item   To mitigate the effect of the non-overlapping organization of the stored patterns we extend the set of possible neurons states to $x_i=0,\pm1$  (i.e. we consider a diluted version of the Hopfield model as done in \cite{Maetschke14}) and propose a slightly modified version of the Hebb's rule (the net effect of this
modification is to have a binary weight matrix with $W_{ij}=p/N$ if $i$ and $j$ belong to the same class and  $W_{ij}=-p/N$ otherwise).
\begin{equation}\label{hebb2}
W_{ij}=\frac{p}{2N}\left(\sum\limits_{\mu=1}^p \xi^{\mu}_i \xi^{\mu }_j  +2-p \right)  
\end{equation}
Moreover we propose an asymmetric version of the updating rule.
During the updating process, we let neurons switch  from $-1$ to $+1$, 
but every time a neuron should be switched from $+1$ to $-1$ we instead eliminate it from the model (i.e. we set the corresponding neuron to zero). This is the reason why we call our version of the Hopfiled model asymmetric. \\
\\
The rationale behind this choice is that Hope4Genes suggests to switch a gene from +1 to -1 if the gene is bringing a misleading signal in the classification problem. This happens for example when a gene in input is over-expressed for purely stochastic reasons. We chose in this case to control the noise by not treating the gene as under-expressed (i.e. set it to -1), but simply neglecting it (i.e. set it to 0) from the model.
Additionally, when a multi-class classification problem is considered, a sample may have no clear association to a specific class being a combination of different classes among which there is a dominant one. Our asymmetric version of the updating rule allows the algorithm to more efficently identify  the dominant class by eliminating from the game genes which are correctly overexpressed in the input (i.e. they belong to the dominant signature and appear with sign +1 in the input) but are wrongly switched off due to the combined pressure of the genes of the other signatures.  Setting them to zero we avoid the amplification of this wrong switch in the following updating steps.  At the same time, due to the 95\% overlap criterion discussed below, if too many of these events happen the algorithm is able to recognize that the input is too ambiguous and returns an "unclassified" result in output.

\item To overcome the presence of unbalanced signatures, which might induce a bias in favor of  the larger signatures, we introduce a cutoff $\Gamma$ on the signature size. Therefore, if the size $N_\mu$ of the signature associated to class $\mu$ is greater than $\Gamma$, $\Gamma$ genes are selected at random from the corresponding signature. When $N_\mu>\Gamma$ all the neurons corresponding to genes deleted from the signature, are set to zero in the input configuration and, due to the updating rule eq.(\ref{agg}), they don't play any further role in the time evolution of the model. To avoid loosing information, we combined the signature reduction with a resampling procedure. Therefore the use of the $\Gamma$ threshold is iterated $M >>1$ times (default $M=100$). For each sample, the class to which the Hopfield model converged in the majority of the $M$ tests is selected as final output of our analysis.\\
\\
The choice of the threshold $\Gamma$ is of crucial importance for the performances of the algorithm. On one side it should be chosen of the same order of the lowest signature size to balance the various classes. Moreover, small $\Gamma$ implies small signatures that are generally less affected by noise (i.e. misclassified genes) than large ones.\\ 
On the other side $\Gamma$ should be chosen as large as possible to maximize the number of error-free patterns to store into the model. Indeed, in a Hopfield model the probability that any bit $i$ is unstable is
$$
P_{error} = \dfrac{1}{2}\left[ 1 - erf\left( \sqrt{\dfrac{N}{2p}}\right) \right], 
$$
where for our application here and in the following $N=N\left(\Gamma\right)$. Given that a pattern is composed of $N$ bits the probability of having an error-free recall of a stored pattern is $\left( 1- P_{error}\right)^{N}$, that we require to be greater than some set value, such as $0.99$. Keeping the binomial expansion and the lowest terms in $P_{error}$, given that this should be small, we get: 
$$
P_{error} = \dfrac{1}{2}\left[ 1 - erf\left( \sqrt{\dfrac{N}{2p}}\right) \right] < \dfrac{0.001}{N}. 
$$
In our case, $p << N$ and thus $erf(x) \approx 1-\dfrac{e^{-x^{2}}}{\sqrt{\pi}x}$. We thus get:
$$
\ln P_{error} \approx -\dfrac{N}{2p} - \ln2 -\dfrac{1}{2}\ln\pi-\dfrac{1}{2}\ln\left( \dfrac{N}{2p}\right) < \ln(0.001) - \ln N, 
$$
from which:
\begin{equation}\label{gamma_rule}
N= N\left(\Gamma\right) > p \ln \left( \dfrac{Np}{0.0002\pi}\right).
\end{equation}
Therefore $\Gamma$ has to be small to avoid noise and to have balanced signatures, but at the same time it has to let $N$ satisfy equation \ref{gamma_rule}. Given that generally $2 \leqslant p \leqslant 6$, according to equation \ref{gamma_rule},  the threshold value of $N\left(\Gamma\right)$ turns out to be in between $70$ and $100$. Since the configurations we are storing in the network are correlated, we decided to select a value of $\Gamma$ so as to have a value of $N$ significantly larger than the above threshold and chose as default value for our analyses $\Gamma = 200$. 
\end{enumerate}

\subsection*{The Hope4Genes algorithm}
Hope4Genes represented in Figure \ref{fig:pipeline} and available at \url{https://github.com/lcan88/Hope4Genes},is composed of five main steps: 

\begin{enumerate}
\item Fixing the synaptic weights of the model
\item Dataset preprocessing
\item Classification of the samples
\item Classification confidence computation
\item Selection of the secondary class
\end{enumerate}
We provide an overview of the algorithmic steps below.

\subsubsection*{1. Fixing the synaptic weights of the model.}
The first input of Hope4Genes is the list of genes that are part of at least one signature. In all the examples that we studied the signatures were composed by $p$ non-overlapping sets of $N_\mu$,
$\mu \in \{1,\ldots,p\}$ genes but in principle our algorithm could address also overlapping signatures. Starting from this first input, we associate to each class $\mu$ a pattern $\xi_\mu$, corresponding to the class template, 
i.e. its components $\xi_\mu^i=1$ if gene $i$ is in the signature of class $\mu$ and $\xi_\mu^i=-1$ otherwise. When the memories are defined the synaptic weights are fixed according to eq.(\ref{hebb2}).\\

The second input is the threshold $\Gamma$ which may be chosen by the user. As discussed above the performances of the algorithm may depend on $\Gamma$. The suggested value that turned to be optimal in all the tested examples is $\Gamma=200$. 

\subsubsection*{2. Dataset preprocessing }
The third input is the gene expression dataset $A$ which is preprocessed, in accordance with \cite{Maetschke14}, as follows:
\begin{itemize}
\item All the genes that are not contained in any signature are filtered out.
\item For each remaining gene $i$, we associate to the corresponding node in the network the initial state 
$$
x_i^0 =sign(log2ratio(A[i])),
$$
where $A[i]$ is the expression value of gene $i$ in the current sample. If $A[i]=0$ (say, because the gene $i$ is not present in the sample) then we set $x_i^0=0$.
\end{itemize}

\subsubsection*{3. Classification of the samples}
We let the initial configuration $x_i^0$ evolve following eq. (\ref{agg}) till it converges to one of the $p$ stored memories. Since some of the neurons are set to zero
during the evolution, in most of the cases it is not possible to find a perfect match of the evolving network state with one of the memories, therefore we assume that
convergence is reached if $95\%$ of the signature neurons match with a class template. The whole classification process is repeated $M$ times with different random choices of the $\Gamma$ genes for all the classes in which $N_\mu>\Gamma$. Finally, each sample is associated to the class to which it converges in the majority of the $M$ replicas. In the algorithm $M$ is fixed to $100$, but the results do not depend on the choice of $M$ as far as $M>>1$.

\subsubsection*{4. Classification confidence computation}
In order to asses the solidity of the predictions made by Hope4Genes we also compute the prediction confidence associated to each sample through the use of the False Discovery Rate (FDR). 

For each sample, the energy of the Hopfiled model (defined by eq.(\ref{energy})) at the initial state $x_i^0$ is used as a proxy of its distance from the stored memories (i.e. the class templates). A null model is then created by reshuffling the expression values of the sample 1000 times, obtaining 1000 random samples that are classified following the procedure described in the section above. Among the 1000 reshuffled samples only those that converged to the same class of the sample under analysis were considered in the null model. In this case, the histogram of their starting energies is constructed and this distribution is used to compute the p-value of the sample under investigation. Given the p-value, the FDR is then computed using the standard Benjamini-Hochberg formula. 

\subsubsection*{5. Selection of the secondary class}
One of the interesting features of our algorithm is that, when dealing with multi-class classification problems it allows to obtain the second best prediction. This prediction is obtained by deleting all the genes belonging to the signature of the convergence class and then repeating all the steps discussed above. The identification of the secondary class may be important in all those cases in which, due to the intrinsic characteristics of the analyzed sample, we expect a certain mixing of different classes as it happens for instance for tumor subtyping.

\bibliography{sample}

\section*{Acknowledgements}

This  work  is  supported  by  the  ”Departments  of  Excellence  2018  -  2022”  Grant  awarded  by  the  Italian  Ministry of Education, University and Research (MIUR) (L.232/2016).

\section*{Author contributions statement}
L.C. and M.C. designed the study. L.C. developed the pipeline software and conduced analysis. L.C. and M.C. wrote the manuscript. M.C. supervised the study.

\section*{Additional information}

The authors declare no competing interests.

\end{document}